\documentclass[floatfix,traditabstract,preprint2]{aa}

\usepackage{natbib, graphicx,rotating}
\begin{document}

\title{A new radiative cooling curve based on an up to date plasma emission code}
\author{K.M. Schure\inst{1}\thanks{email: K.M.Schure@phys.uu.nl}
\and D. Kosenko \inst{1}
\and J.S. Kaastra \inst{1,2}
\and R. Keppens \inst{1,3,4}
\and J. Vink \inst{1}}
\institute{Astronomical Institute, University of Utrecht,
             Postbus 80000, NL-3508 TA Utrecht; K.M.Schure@phys.uu.nl
\and SRON, Utrecht, The Netherlands
\and Centre for Plasma Astrophysics, K.U. Leuven, Belgium
\and FOM-Institute for Plasma Physics ``Rijnhuizen'', Nieuwegein, The Netherlands}

\date{ Received \ldots / Accepted \ldots}

\abstract{
This work presents a new plasma cooling curve that is calculated using the SPEX package. We compare our cooling rates to those in previous works, and implement the new cooling function in the grid-adaptive framework `AMRVAC'. Contributions to the cooling rate by the individual elements are given, to allow for the creation of cooling curves tailored to specific abundance requirements.
In some situations, it is important to be able to include radiative losses in the hydrodynamics. The enhanced compression ratio can trigger instabilities (such as the Vishniac thin-shell instability) that would otherwise be absent. For gas with temperatures below $10^4$~K, the cooling time becomes very long and does not affect the gas on the timescales that are generally of interest for hydrodynamical simulations of circumstellar plasmas. However, above this temperature, a significant fraction of the elements is ionised, and the cooling rate increases by a factor 1000 relative to lower temperature plasmas. 
}


\authorrunning{Schure et al.}
\titlerunning{SPEX cooling curve}
\maketitle

\section{Introduction}

Optically thin plasmas can cool very efficiently, especially on timescales relevant to circumstellar evolution. A correct description of cooling is therefore important for simulations of e.g., stellar winds. The dominant cooling mode for plasmas within the temperature range of $10^4-10^7$~K is by metal line transitions. In many cases, the plasma can be assumed to be in collisional ionisation equilibrium (CIE). A predefined cooling curve can then be used to account for radiative losses, where cooling rates depend only on the temperature of the plasma. In the calculation of the cooling curve, we can adjust the abundances and try to incorporate as many line transitions as are reliably known, to obtain a straightforward and efficient way of treating radiative cooling. 
Well-documented cooling curves used in HD and MHD computations are those for example developed by \citet{1972DalgarnoMcCray},\citet{1977RaymondSmith}, \citet{1981MacDonaldBailey},\citet{1993SutherlandDopita}, and \citet{2008Smithetal}. Other approaches, such as that of \cite{2002MellemaLundqvist}, include time- and ion dependent cooling, with CSM abundances that depend on the evolutionary stage of the star, but treat the hydrodynamics only in one dimension.

We use the SPEX package \citep{1996Kaastraetal} to calculate a new cooling curve, using up-to-date transition lines to calculate radiative losses for a plasma. This code is used as a spectral analysis code tailored to EUV and X-ray observations, in which energy band it is one of the most complete packages currently available. Because it includes a very complete prescription of line emission, cooling rates predicted by SPEX are higher than those of cooling curves available until now. 

We implement the new cooling curve, as deduced from the SPEX CIE code, in the AMRVAC framework~\citep{2003Keppensetal}, of which we use the hydrodynamics to demonstrate the influence of radiative losses on the gas dynamics. As an example, we use this code with the new cooling prescription to simulate the circumstellar medium of a massive star. 
Plasmas typical of this circumstellar medium are at fairly high temperatures and are mostly optically thin. Therefore, they can cool very efficiently. 

The cooling at a certain point in the simulated plasma can be calculated from the local temperature and density in combination with the prescribed cooling curve. Different cooling curves can be chosen, which depend on metallicity and/or ionisation fraction (for low temperatures). We provide cooling rates for the individual elements, so that the cooling curve can be tailored to fit the required abundance ratios. 

In Sect.~\ref{sec:spex}, the SPEX code, the abundances, and the new cooling curve will be presented and compared to existing cooling curves. We show that the SPEX curve causes more cooling for temperatures from $10^4$ to $10^6$~K, a temperature range that is very important in the evolution of the circumstellar as well as the interstellar medium. 

In Sect.~\ref{sec:cie}, we discuss the range of validity of the CIE assumption and discuss the implications of photoionisation and shock heating on the cooling rates of the plasma.

Cooling at lower temperatures cannot be accurately calculated with SPEX, because of incomplete data for $\lambda>2000$~\AA. 
In Sect.~\ref{sec:DMcooling}, we briefly summarise our treatment of lower temperature plasmas, which follows \citet{1972DalgarnoMcCray}. The cooling rates of \citet{2008Smithetal}, provide a detailed representation of especially the low-temperature ($T <10^4$~K) cooling of plasmas. We select our low-temperature ionisation fraction to ensure the optimal correspondence with their cooling rates.

In Sect.~\ref{sec:test}, we show the effect that cooling has on the dynamics of the plasma, by using the SPEX cooling curve as implemented in the AMRVAC code. Because the cooling adds to the resolution requirements of the hydro-simulations, the grid-adaptive method used by AMRVAC maximises the performance while reaching sufficient resolution to enhance the higher density contrast and allow for instabilities to develop in the cooling shell. 
In Sect.~\ref{sec:conclusion}, we summarise the important points and outline the major differences from earlier cooling curves.

\section{SPEX cooling curve} 
\label{sec:spex}

To generate the cooling curve we use the {\sc SPEX}\footnote{http://www.sron.nl/spex} package, version 2.00.11 \citep{2000KaastraMewe}. This code is tailored to fit spectra in the EUV and X-ray regime and can be used to calculate the luminosity of a given emission measure for plasmas at different temperatures and different choices of electron and ion temperatures. For our purpose, we use SPEX to calculate the emissivity of a plasma in collisional ionisation equilibrium (CIE). The code is based on the mekal line-emission model \citep{1995Meweetal}, but many updates have since been made. The 15 elements presently included are H, He, C, N, O, Ne, Na, Mg, Al, Si, S, Ar, Ca, Fe, and Ni. 

We calculate spectra for a grid of temperatures ranging from $10^{3.8}$ to $10^{8.16}$~K, with steps of $\log T=0.04$~K. Solar abundances are assumed, using the abundance table presented in \citet{1989AndersGrevesse}, which is summarised here in Table~\ref{table:AGabun} for elements that are taken into account. The emissivity $\Lambda_N$~(erg~s$^{-1}$~cm$^{3}$, normalised to $n_{\rm e} n_{\rm H}V=1$~cm$^{-3}$) is calculated by integrating the spectrum over an energy range from $10^{-1}$~eV to 1~MeV. The SPEX package is currently one of the most complete spectral packages, and it is especially complete for temperatures higher than $\sim 10^{4.86}$~K (see Sect.~\ref{sec:DMcooling}). 

\begin{figure}[!htbp]
     \centering
  \resizebox{\hsize}{!}{

     \includegraphics{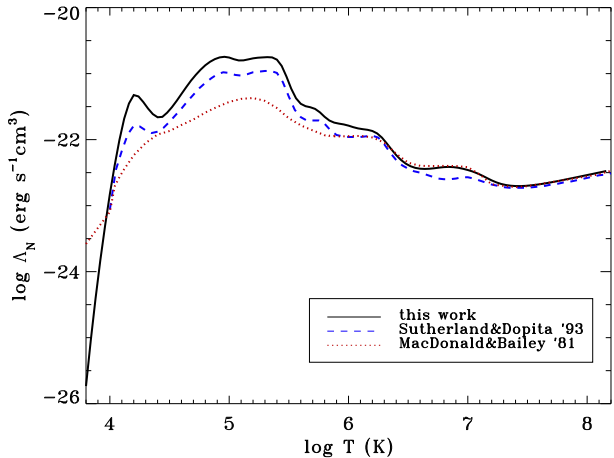}
   }
      \caption[ ] {Cooling curves compared: The higher cooling rates calculated with SPEX are mainly due to a more complete coverage of the line transitions, including Fe L and EUV lines.
      \label{fig:curves}}
      \end{figure}

The cooling curve is tabulated in Table~\ref{table:KSsolar}, and in Fig.~\ref{fig:curves} it is plotted against the CIE cooling curve that was calculated by \citet{1993SutherlandDopita}. Compared to their curve, and other CIE cooling curves currently available, we find that the cooling efficiency calculated by SPEX is higher. We attribute this difference to SPEX taking into account more line transitions in both X-ray and EUV than previous studies (5466 lines in SPEX compared to 1683 in \citet{1993SutherlandDopita}). Additionally, though \citet{1993SutherlandDopita} use the same values of Solar abundance ratios, they use a slightly different ionisation balance, which they calculate with their MAPPINGS~II code. SPEX uses the ionisation balance from \citet{1985ArnaudRothenflug} for most elements, apart from the one from \citet{1992ArnaudRaymond} for iron \citep{1995Meweetal}. A more complete list of Fe L complex lines is also included, based on the HULLAC atomic physics package \citep{1977Klapischetal}, causing higher cooling efficiency around $10^7$~K. The differences in the treatment of Fe are shown in \citet{1995Meweetal}. 

Additionally, we plot the cooling curve by \citet{1981MacDonaldBailey} (hereafter also referred to as MB81) in Fig.~\ref{fig:curves}. We used this for comparison in our hydro-simulations in Sect.~\ref{sec:test}. This cooling curve is a composite of cooling rates by \citet{1976Raymondetal} for $10^6<T\le 10^8$~K and by \citet{1976ShapiroMoore} for $10^4<T<10^6$~K. Below $10^4$~K, extrapolation of the cooling rates is used, towards zero at 100~K. Their treatment of cooling between $10^4-10^6$~K is a non-equilibrium approach, applicable to a low density plasma that 
is overionised, as a result of rapid cooling ($t_{\rm cool} < t_{\rm recombination}$).
Time-dependent cooling, starting from ionisation equilibrium at $10^6$~K, gives rise to lower cooling rates, as discussed in Sect.~\ref{sec:cie}.  MB81 base their cooling rates on the inclusion of 12 elements (H, He, C, N, O, Ne, Mg, Si, S, Ca, Fe, Ni), and use slightly different abundances, which are not however responsible for the major differences from our cooling rates.

To apply the cooling curve in the AMRVAC code, we renormalise the cooling rates calculated with SPEX by multiplying by the electron to proton ratio ($n_{\rm}e/n_{\rm H}$). We thus obtain a value for $\Lambda_{\rm hd}$ that we can use in our hydrodynamic application: 
\begin{eqnarray}
\Lambda_{\rm hd}=\frac{n_{\rm e}}{n_{\rm H}}\Lambda_N.
\end{eqnarray} 
The ratio $n_{\rm e}/n_{\rm H}$ is a temperature-dependent parameter that is also calculated by SPEX and is given in the last column of Table~\ref{table:KSsolar}. The cooling rate of the plasma is given by
\begin{eqnarray}
L = \int n_{\rm H} n_{\rm H} \Lambda_{\rm hd}(T) dV \quad({\rm erg~s}^{-1}).
\end{eqnarray}

\begin{table}[!htbp]
\centering                 
\caption{\textrm{Solar abundances from \citet{1989AndersGrevesse} }  }         
\label{table:AGabun}        
\begin{tabular}{c  c | c c }       
\hline\hline 
element& $\log (n/n_{\rm H})$ & element &$\log (n/n_{\rm H})$\\
\hline
He & -1.01 & Al & -5.53\\
C & -3.44 & Si & -4.45\\
N & -3.95 & S & -4.79\\
O & -3.07& Ar & -5.44\\
Ne & -3.91 & Ca & -5.64\\
Na & -5.67 & Fe & -4.33\\
Mg & -4.42 & Ni & -5.75\\

\hline
\end{tabular}
\end{table}

\begin{figure}[!htbp]
     \centering
   \includegraphics[width=\hsize]{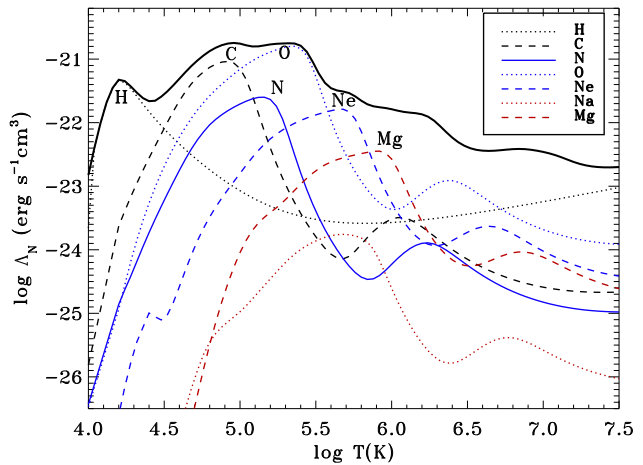}
   \includegraphics[width=\hsize]{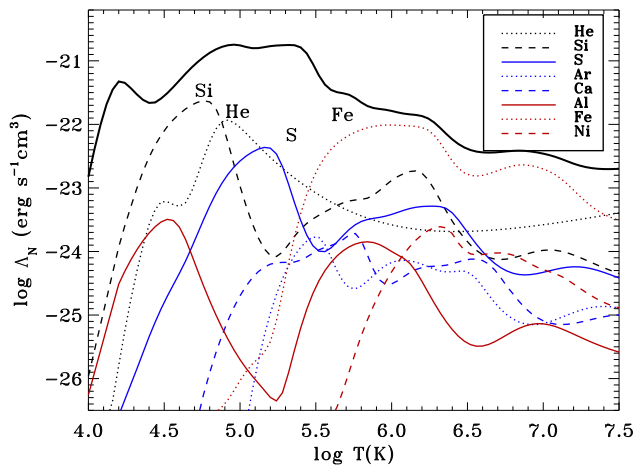}
      \caption[ ] {Contributions of different elements to the cooling curve are given. Each of the plots shows a different set of elements. Important peaks are labelled with the name of the element. The total cooling curve (black solid line) is an addition of the individual elemental contributions. 
      \label{fig:lambda_abun}}
      \end{figure}

To allow the cooling curve to be adapted for different abundances, we list the contributions of the different elements to the cooling rates, as calculated with the SPEX code. This is tabulated in Table~\ref{table:KSabun} (online only). The cooling rates are given again for solar abundances \citep{1989AndersGrevesse}. The contributions of the individual elements are plotted in Fig.~\ref{fig:lambda_abun} to illustrate the relative importance of the different elements to certain temperature ranges. 
The cooling rates are given for normalised values of $n_{\rm H} n_{\rm e}=1$, and should be multiplied by $n_{\rm e}/n_{\rm H}$ when used in a hydrocode.

For different metallicities, we can easily construct new cooling curves from Table~\ref{table:KSabun} (online only), in which the cooling contributions and ionisation fractions of the elements are given separately. Cooling rates could then be calculated by summing the contributions of the different elements, multiplied by the appropriate abundance ratio and ionisation fraction
\begin{eqnarray}
\Lambda_N(T)=\sum_i \frac{n_i}{n_i ({\rm solar})} \Lambda_N(X_i,T),
\end{eqnarray} 
where $n_i$ is the abundance ratio (relative to H) of element $X_i$.
In this way, new cooling curves can be created ad hoc when needed for simulations of plasmas with different metallicities.

\section{Assumptions and validity of CIE cooling rates}
\label{sec:cie}

Collisional ionisation equilibrium is a valid assumption if an optically thin plasma is dominated by collisional processes and the cooling timescale is longer than either the ionisation or recombination timescales. Deviations from CIE can occur if there is additional photo\-ionisation or there is non-equilibrium ionisation (NEI), i.e., the recombination or ionisation timescales are longer than the cooling/heating timescale. These circumstances will produce an overionised plasma if photo\-ionisation is present or if the cooling timescale is shorter than the recombination timescale. In young objects, such as supernova remnants, the heating timescale is shorter than the ionisation timescale, giving rise to an underionised plasma.

NEI requires a time-dependent treatment of the cooling and is not easily implemented in standard hydrodynamical codes, but some specialised one-dimensional hydrodynamical codes for SNRs do exist \citep[e.g.,][]{2003Badenesetal,2004Kosenkoetal}. On the other hand, photo\-ionisation requires additional ionisation dependence on the photon-field, resulting in a higher ionisation than CIE for the same plasma temperature. The efficiency and reach of a photo\-ionising source depends on both its spectrum and the ionisation parameter $\xi = \frac{L}{nR^2}$, where $R$ is the distance to the source. Depending on whether the elements are more or less effective coolants in their additionally ionised state, the cooling rates may be higher or lower, respectively. 

Studies of the cooling rates of over\-ionised plasmas, relevant to photo\-ionised plasmas or the late radiative stages of supernova remnants, have been performed by various authors \citep[e.g.,][]{1973Kafatos, 1976ShapiroMoore, 1993SutherlandDopita}. The cooling rates in all of these cases are lower than that of a CIE cooling curve.  For comparison, we plot the cooling curve by MB81 in Fig.~\ref{fig:curves}. This curve uses the time-dependent cooling curve by \citet{1976ShapiroMoore} for the temperature range $4 < \log T < 6$. The gas is initially assumed to be in ionisation equilibrium at $T=10^6$~K, and cools time-dependently from there. Since the cooling timescale is shorter than the recombination timescale, the gas becomes over\-ionised, leading to suppressed cooling rates in this regime. Similarly, in intergalactic plasmas where photo\-ionisation is important, cooling rates tend to be suppressed  \citep{2009Wiersmaetal}. Alternatively, cooling rates may be much higher when a neutral gas is shock-heated, due to the initial ionisation of H and He \citep{1985CoxRaymond}. 
 Unless full radiative transfer is taken into account, we therefore find CIE to be a reasonable assumption to use in circumstellar environments and other low density plasmas. Only when photo\-ionisation is significant, which depends on the temperature of the star and the distance to the source, or when strong shocks with velocities higher than about 150~km~s$^{-1}$ are present, deviation from CIE can become important. At higher shock speeds, the gas is heated to temperatures higher than $\log T \approx 5.7$, at which cooling by H and He lines of C, N, and O, becomes suppressed. Even in our relatively strong shock between two subsequent wind-phases as presented in Sect.~\ref{sec:test} however, the typical speed at the forward shock is less than 150~km~s$^{-1}$ and the temperature is typically around $10^5$~K.

Since recombination, collisional ionisation, and cooling have similar density dependences, the cooling rates that we calculate are independent of density, as long as the gas can be regarded as optically thin.

\section{Low temperature cooling}
\label{sec:DMcooling}
The SPEX package is tailored to the X-ray regime, and includes continuum but no line emission for wavelengths longer than $\lambda>2000$~\AA. Therefore, it may underestimate cooling for temperatures below $T=hc/\lambda k \sim10^{4.86}$~K. At temperatures below $10^4$~K, cooling is much less efficient, and in the case of the typical CSM that we are interested in, most cooling takes place above this threshold. 

For low temperatures (T$\ <10^4$~K), we use the cooling rates from a number of reactions as given by \citet{1972DalgarnoMcCray}. 
In this regime, fractional ionisation and molecular line transitions can play an important role. We can set the ionisation fraction $f_{i}=n_{\rm e}/n_{\rm H}$, which is a free parameter, to accomodate for different circumstances in the circumstellar medium (CSM). This fraction strongly depends on specific conditions concerning shocks, radiation, and the history of the plasma. The cooling rates are included in Table~\ref{table:DMsolar} (online only), and below we repeat the equations that we use to calculate them. The values for an ionisation fraction of $10^{-4}$ correspond most closely to the values that \citet{2008Smithetal} find for a Solar abundance plasma. 

The elements that are taken into account for low-temperature cooling are O, C, N, Si, Fe, Ne, and S. The cooling is provided partly by excitation of singly charged ions with the thermal electrons. The electron-collision transitions that are taken into account are given by the following equations (with the temperature T in Kelvin) \citep{1972DalgarnoMcCray}: 
\begin{eqnarray}
\label{eq:Ce}
\Lambda_{\rm e}({\rm C^+})&=&\left(\frac{n_i}{n_{\rm H}} f_i \right)T^{-1/2}\left[ 7.9\times10^{-20}e^{-92/T}\right. \\\nonumber
       &+& \left. 3.0 \times 10^{-17} e^{-61900/T}\right]\\
\Lambda_{\rm e}({\rm Si^+})&=&\left(\frac{n_i}{n_{\rm H}} f_i \right)T^{-1/2}\left[ 1.9\times10^{-18}e^{-413/T}\right. \\\nonumber
       &+& \left. 3.0\times 10^{-17}e^{63600/T}\right]\\
\Lambda_{\rm e}({\rm Fe^+})&=&\left(\frac{n_i}{n_{\rm H}} f_i \right)T^{-1/2}\left[4.8\times 10^{-18}e^{2694/T}\right.  \\\nonumber  
&+&1.1\times10^{-18}\left(e^{-554/T}+1.3e^{-961/T}\right)\\\nonumber
       &+& \left. 7.8\times 10^{-18}e^{3496/T}\right]\\
\Lambda_{\rm e}({\rm O})&=&\left(\frac{n_i}{n_{\rm H}} f_i \right)1.74\times10^{-24}T^{1/2} \\\nonumber
                                    &\times&[(1-7.6T^{-1/2})e^{-228/T}\\\nonumber
                                   &+&0.38(1-7.7T^{-1/2})e^{-326/T}]\\\nonumber
&+&9.4\times10^{-23}T^{1/2}e^{-22700/T}\\
\Lambda_{\rm e}({\rm O^+})&=&\left(\frac{n_i}{n_{\rm H}} f_i \right)1.5\times10^{-17}T^{-1/2}e^{-38600/T}\\
\Lambda_{\rm e}({\rm N})&=&\left(\frac{n_i}{n_{\rm H}} f_i \right)8.2\times10^{-22}T^{1/2}\\\nonumber
                                    &\times&(1-2.7\times10^{-9}T^2)e^{-27700/T}\\
\Lambda_{\rm e}({\rm S^+})&=&\left(\frac{n_i}{n_{\rm H}} f_i \right)8.4\times10^{-18}T^{-1/2}e^{-21400/T}.\\
\end{eqnarray}

For low fractional ionisation, collisions with neutral hydrogen can add substantially to the cooling. The following transitions are taken into account:
\begin{eqnarray}              
\Lambda_{\rm H}({\rm Si^+})&=&\left(\frac{n_i}{n_{\rm H}}\right)7.4\times10^{-23}T^{-1/2}e^{-413/T}\\
\label{eq:FeH}
\Lambda_{\rm H}({\rm Fe^+})&=&\left(\frac{n_i}{n_{\rm H}}\right)1.1\times10^{-22}\times\\\nonumber
                             &&\left[e^{-554/T}+1.4e^{-961/T}\right]
\end{eqnarray}
The total cooling that we take into account  for $T < 10^4$~K is a summation of contributions from Eq.~\ref{eq:Ce} to \ref{eq:FeH}, plus contributions from  collisions between electrons and hydrogen atoms as tabulated in Table~2 in \citet{1972DalgarnoMcCray}.

\section{Application: Test problem}
\label{sec:test}

We use the AMRVAC code with the new cooling prescription to simulate the circumstellar medium (CSM) of a massive star. The adaptive mesh refinement (AMR) ensures a computationally efficient approach to problems that have localised needs of high resolution, but where lower resolution suffices in substantial parts of the grid. The standard Euler equations are solved in a conservative way, where we apply an HLLC solver \citep{1999Toro} to capture more accurately the contact discontinuity.

We implement the stages of the evolution of a massive star in which a red supergiant (RSG) wind is followed by the much faster Wolf-Rayet (WR) wind. A RSG wind is typically slow, cool, and dense because of the low escape velocity from the surface of the giant star. The WR wind, however, is fast and hot, originating from the compact stellar core. The WR wind therefore sweeps up the RSG wind into a shell \citep{1995GGSMacLow,1996GGSetal_II,2007Dwarkadas}, hereafter called the WR shell, whose thickness and stability is strongly affected by the inclusion of radiative cooling. 

We set up the simulation in 2.5D on a spherical grid in a meridional plane over a full angle $\pi$, using reflective boundary conditions and symmetry around the $\phi$-axis. The radial extent is about 3~pc and we use a base resolution of $60\times48$ gridcells with 4-6 refinement levels. For each refinement level, the resolution is doubled. The values that we use for the wind parameters are the following: $\dot M_{\rm RSG}=1.54\times 10^{-5} M_\odot$~yr$^{-1}$, $v_{\rm RSG}=4.7$~km~s$^{-1}$, $T_{\rm RSG}=10^3$~K, $\dot M_{\rm WR}=9.7\times 10^{-6} M_\odot$~yr$^{-1}$, $v_{\rm WR}=1.7\times10^3$~km~s$^{-1}$, $T_{\rm RSG}=10^4$~K. 

For the RSG phase, 4 refinement levels are used, and for the WR phase, we use 4-7 refinement levels and check for differences in the development of the instabilities and the resolution dependence of the radiative losses. The maximum resolution corresponds to $1.56\times10^{15}$~cm $\times\ 0.0156^\circ$.

In general, all plasmas are prone to radiative losses. Depending on the circumstances however, these can induce significant changes in the hydrodynamics, or have only minor quantitative effects.
In the situation that we simulate, without cooling the maximum compression reached in the shell is a factor of four; the value expected for a strong shock in a mono-atomic gas with an adiabatic index of $\gamma=5/3$. With cooling, the shell can become much more compressed and, as a result, much thinner. 
When the compression ratio exceeds a factor of about $21$, the shell can become subject to the Vishniac thin shell instability \citep{1983Vishniac, 1987RyuVishniac,1993MacLowNorman}. 

The Vishniac thin shell instability, which actually is an overstability, occurs when a thin shell is perturbed while being driven by a high pressure region from the inside and encountering ram pressure on the outside. The high pressure from the WR wind acts in a perpendicular direction to the shell, whereas the ram pressure acts in a parallel direction to the velocity. Whereas the velocity is radial, the interior pressure in the case of a perturbed shell can create a flow along the shell surface, driving matter from advanced regions to regions lagging behind. The more tenuous regions then become more susceptible to deceleration by ram pressure, reversing the cycle. 

Above a temperature of $~10^5$~K, the plasma also becomes susceptible to the thermal instability. This occurs for shock velocities above $~150$~km~s$^{-1}$, when $({\rm d}\ln \Lambda)/({\rm d}\ln T) < 2$ \citep[e.g.,][]{1965Field, 1986Bertschinger, 1993SutherlandDopita}. For typical circumstellar wind parameters, the plasma temperature in the cooling shell remains below this regime.

The cooling of the plasma is calculated at each timestep and the thermal energy is correspondingly updated. The temperature of the plasma is first determined. The corresponding cooling rate is then found by interpolating between the appropriate values in the cooling curve. The energy loss rate per unit volume is calculated as follows:
\begin{eqnarray}
\frac{{\rm d}e}{{\rm d}t}=-\Lambda_{\rm hd}n_{\rm H}^2.
\end{eqnarray}
The energy losses are subtracted from the thermal energy only, preventing numerical precision problems from producing negative energy.

Although the cooling rates are higher when we use the SPEX-calculated cooling curve, in both cases the thickness of the WR shell is not so much limited by the cooling curve used, as by the adopted numerical resolution. It is important however, that the shell becomes affected by instabilities in the presence of cooling, whereas it remains stable in the absence of cooling. We compare the differences for runs with different resolution and different cooling prescription. For comparison, we use the cooling curve as prescribed in \citet{1981MacDonaldBailey}, since this is the cooling curve that is used in the ZEUS3D v3.4 code, as used by some of the authors before, for simulations of a WR wind in \citet{2008Schureetal}.

\begin{figure}[!htbp]
\centering

  \resizebox{\hsize}{!}{
    \includegraphics{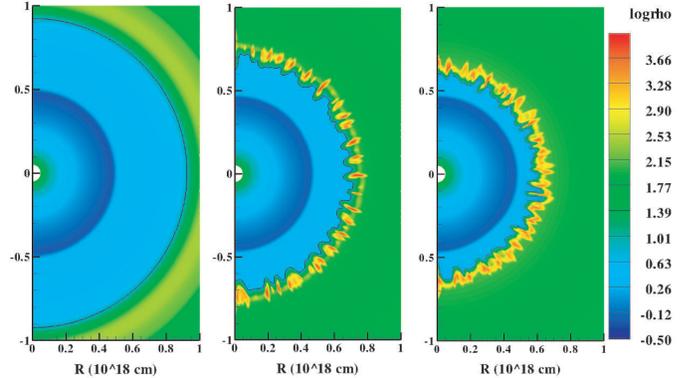}}
      \caption{WR wind colliding with RSG wind. The density of the CSM is shown about 1750 years after the initiation of the WR wind, showing the early development of the Vishniac thin shell instability. The left panel shows the CSM when no radiative losses are taken into account. The middle panel shows the case where the MB81 cooling prescription is applied, and the right panel shows the CSM with the cooling prescription as presented in this work. The black line indicates the contact discontinuity. The number of refinement levels used for the resolution is 5, i.~e., $6.25 \times 10^{15}$~cm $\times\ 0.0625^\circ$.
      \label{fig:rhonombks}}
\end{figure}

Figure~\ref{fig:rhonombks} shows the density of the circumstellar medium after the WR-phase has just set in for about 1750 years, and shows the early developments of the Vishniac thin shell instability. The fast wind from the Wolf-Rayet star has swept up the red supergiant wind of the previous evolutionary stage and four regions can be identified. From inside (R=0) out: the free streaming fast WR wind, the shocked WR wind. The black line indicates the contact discontinuity, behind which the shocked, and farther out the free streaming, RSG wind can be found. Only part of the grid is shown to enlarge the interesting region.

The left panel shows the CSM for the colliding winds in the absence of cooling. The shell is smooth and thick, with a maximum compression ratio that is similar to that of a strong shock, i.~e., a factor 4. The forward shock is partly outside the plotted region of the grid and extends to a radius of about $1.2 \times 10^{18}$~cm. The middle and right panels illustrate results when cooling is present, for the cooling rate given by \citet{1981MacDonaldBailey} in the middle panel, and for the cooling rate calculated with the {\sc SPEX} code in the right panel. The instability is induced by random density perturbations of the 1\% level, which may induce little differences between runs. If we do not include the initial perturbations, the code retains the original spherical symmetry to high accuracy, and the instabilities do not arise until very late. 
The extrapolation of the MB81 cooling curve towards lower temperatures results in higher cooling rates below $10^4$~K. This gives rise to more cooling in the compressed RSG shell and a more patchy structure as can be seen in the middle panel of Fig.~\ref{fig:rhonombks}. For temperatures above $10^4$~K, cooling rates are higher for the {\sc SPEX} curve. This applies to the region of shocked WR wind material and results in this area being more compressed in the right panel. Consequently, the WR shell in the middle panel extends a little farther out. 

\begin{figure}[!htbp]
\centering
  \resizebox{\hsize}{!}{
      \includegraphics{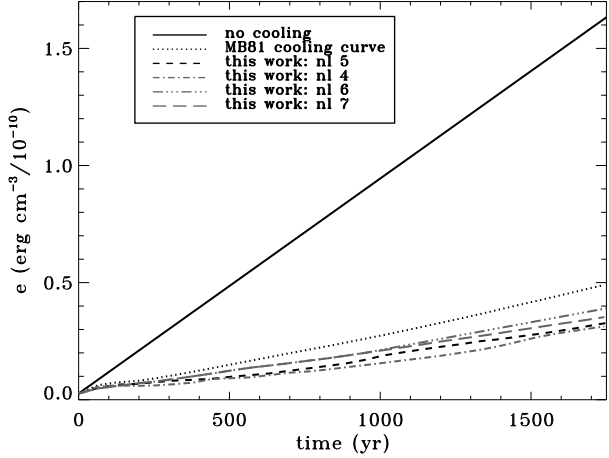}}
      \caption[ ] {Thermal energy in the stellar winds. The black solid line shows the energy when no cooling is taken into account. The black dotted curve shows the energy when the MB81 cooling rates are used, and the black dashed curve shows the system for the cooling rates presented in this work. The gray lines show the resolution dependence of the thermal energy, where $nl$ indicates the number of refinement levels used. 
            \label{fig:ethnombks}}
 \end{figure}

Total energy losses are higher with our new cooling prescription. The comparison is made in Fig.~\ref{fig:ethnombks}. The thermal energy per unit volume is plotted for the different cases in Fig.~\ref{fig:ethnombks}. The inclusion of cooling reduces the thermal energy substantially. We note again that the initial random perturbation will influence the later evolution, explaining some of the inter-resolution differences. However, the key point is the systematic reduction between the SPEX versus MB81 cooling prescriptions.

\begin{figure}[!htbp]
\centering
  \resizebox{\hsize}{!}{
    \includegraphics{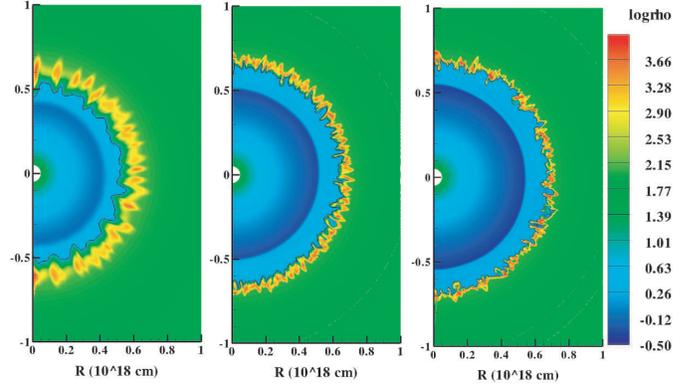}}
      \caption{The Vishniac thin shell instability depends on numerical resolution. At high resolution, the shell becomes thinner and thus the wavelength shorter. The different panels correspond to refinement levels of 4, 6, and 7, corresponding to maximum resolution of $1.25 \times 10^{16}$~cm $\times\ 0.125^\circ$,  $3.125\times10^{15}$~cm $\times\ 0.03125^\circ$, and $1.56\times 10^{15}$~cm $\times\ 0.0156^\circ$. The simulation with 5 refinement levels is shown in Fig.~\ref{fig:rhonombks}.
 \label{fig:rho_ks456}}
\end{figure}

The wavelength of the Vishniac thin shell instability depends on the thickness of the shell. In Fig.~\ref{fig:rho_ks456}, we show the effect of resolution on the development of the instabilities with the new cooling curve. For higher resolution, the shell becomes thinner, and as a result the wavelength of the instability shortens. Although the WR shell is not resolved in terms of the dynamics, we see that for thermal losses, convergence seems to be reached for the resolutions used. Figure~\ref{fig:ethnombks} shows the resolution dependence of the thermal losses for the different cooling prescriptions and different resolutions. 

\begin{figure}[!htbp]
\centering
  \resizebox{\hsize}{!}{
   \includegraphics{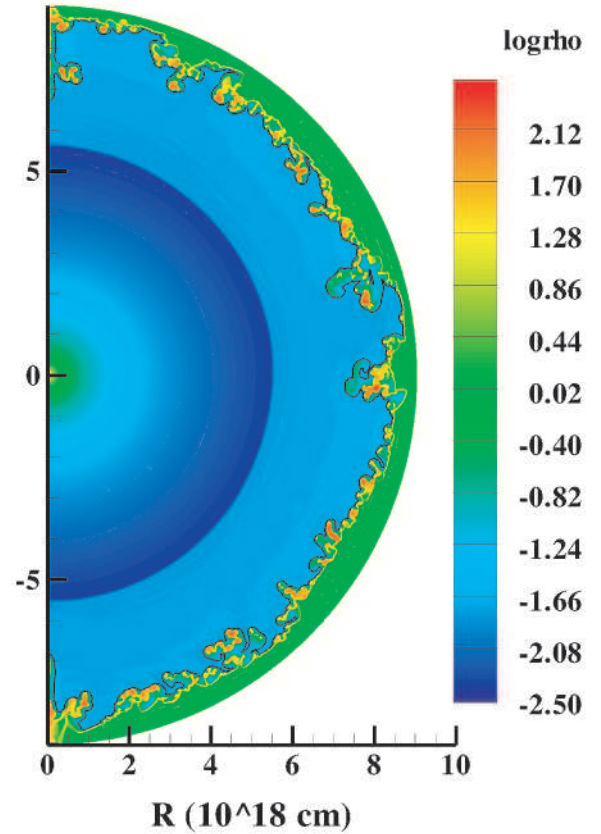}}
      \caption{Unstable WR shell at a time of about 22,700 yrs after its transition to a WR star. The shell becomes highly irregular due to the Vishniac thin shell instability. The black line indicates the contact discontinuity. This simulation used 6 refinement levels for the resolution.
      \label{fig:rhowrks130}}
\end{figure}

A typical WR phase lasts for about $10^5$ years. In Fig.~\ref{fig:rhowrks130}, we show the development of the WR shell later in the evolutionary stage, at which time the WR phase has lasted for about 22,700 yr. It is clear that the presence of cooling majorly alters the CSM evolution, by losing energy by radiation, and by the irregular shell that develops as a result of this. At later times, shells from earlier evolutionary phases may intervene with the CSM evolution and simulations need to take these phases into account \citep[see e.g.,][]{2007Dwarkadas,2007vanMarleetal}. 
      
\section{Conclusion and discussion}
\label{sec:conclusion}

SPEX is a very complete package that calculates the emissivity of a plasma at a certain temperature, taking into account the abundances. We calculate the radiative losses for a plasma in CIE, and compare these with the curve by \citet{1993SutherlandDopita}, which was calculated with the same basic assumptions. We find that cooling rates calculated by SPEX tend to be higher than those of \citet{1993SutherlandDopita}. This is not surprising, since more lines have since been added and the SPEX package is, to our knowledge, the most complete package for X-rays and EUV. For low temperatures, we reuse the cooling prescription by \citet{1972DalgarnoMcCray}. Using that, and the SPEX results, we provide a recommended cooling curve for Solar abundances\footnote{http://www.phys.uu.nl/$\sim$schure/cooling}, which can be adjusted to different ionisation fractions at low temperatures. Additionally, changes can be decided for each cooling curve according to specific requirements on the abundances, by using the separate cooling rates of the individual elements.

In our application, we calculate the CSM of a massive star that experiences a RSG phase followed by a WR phase. The colliding winds form a shell that, in the presence of cooling, becomes very thin and susceptible to the Vishniac thin shell instability. Taking into account the cooling of the plasma is very important when accurately simulating systems such as these.

\begin{table}[!htbp]
\centering                 
\caption{\textrm{Cooling curve for solar metallicity} }           
\label{table:KSsolar}        
\begin{tabular}{c  c c c}       
\hline\hline               
log~$T$& log $\Lambda_N$ & log $\Lambda_{\rm hd}$ & $n_{\rm e}/n_{\rm H}$\\
(K) & (erg~s$^{-1}$~cm$^{3}$) & (erg~s$^{-1}$~cm$^{3}$) & \\
\hline
3.80 & -25.7331 & -30.6104 & 1.3264E-05\\
3.84 & -25.0383 & -29.4107 & 4.2428E-05\\
3.88 & -24.4059 & -28.4601 & 8.8276E-05\\
3.92 & -23.8288 & -27.5743 & 1.7967E-04\\
3.96 & -23.3027 & -26.3766 & 8.4362E-04\\
4.00 & -22.8242 & -25.2890 & 3.4295E-03\\
4.04 & -22.3917 & -24.2684 & 1.3283E-02\\
4.08 & -22.0067 & -23.3834 & 4.2008E-02\\
4.12 & -21.6818 & -22.5977 & 1.2138E-01\\
4.16 & -21.4529 & -21.9689 & 3.0481E-01\\
4.20 & -21.3246 & -21.5972 & 5.3386E-01\\
4.24 & -21.3459 & -21.4615 & 7.6622E-01\\
4.28 & -21.4305 & -21.4789 & 8.9459E-01\\
4.32 & -21.5293 & -21.5497 & 9.5414E-01\\
4.36 & -21.6138 & -21.6211 & 9.8342E-01\\
4.40 & -21.6615 & -21.6595 & 1.0046\\
4.44 & -21.6551 & -21.6426 & 1.0291\\
4.48 & -21.5919 & -21.5688 & 1.0547\\
4.52 & -21.5092 & -21.4771 & 1.0767\\
4.56 & -21.4124 & -21.3755 & 1.0888\\
4.60 & -21.3085 & -21.2693 & 1.0945\\
4.64 & -21.2047 & -21.1644 & 1.0972\\
4.68 & -21.1067 & -21.0658 & 1.0988\\
4.72 & -21.0194 & -20.9778 & 1.1004\\
4.76 & -20.9413 & -20.8986 & 1.1034\\
4.80 & -20.8735 & -20.8281 & 1.1102\\
4.84 & -20.8205 & -20.7700 & 1.1233\\
4.88 & -20.7805 & -20.7223 & 1.1433\\
4.92 & -20.7547 & -20.6888 & 1.1638\\
4.96 & -20.7455 & -20.6739 & 1.1791\\
5.00 & -20.7565 & -20.6815 & 1.1885\\
5.04 & -20.7820 & -20.7051 & 1.1937\\
5.08 & -20.8008 & -20.7229 & 1.1966\\
5.12 & -20.7994 & -20.7208 & 1.1983\\
5.16 & -20.7847 & -20.7058 & 1.1993\\
5.20 & -20.7687 & -20.6896 & 1.1999\\
5.24 & -20.7590 & -20.6797 & 1.2004\\
5.28 & -20.7544 & -20.6749 & 1.2008\\
5.32 & -20.7505 & -20.6709 & 1.2012\\
5.36 & -20.7545 & -20.6748 & 1.2015\\
5.40 & -20.7888 & -20.7089 & 1.2020\\
5.44 & -20.8832 & -20.8031 & 1.2025\\
5.48 & -21.0450 & -20.9647 & 1.2030\\
5.52 & -21.2286 & -21.1482 & 1.2035\\
5.56 & -21.3737 & -21.2932 & 1.2037\\
5.60 & -21.4573 & -21.3767 & 1.2039\\
5.64 & -21.4935 & -21.4129 & 1.2040\\
5.68 & -21.5098 & -21.4291 & 1.2041\\
5.72 & -21.5345 & -21.4538 & 1.2042\\
5.76 & -21.5863 & -21.5055 & 1.2044\\
5.80 & -21.6548 & -21.5740 & 1.2045\\
5.84 & -21.7108 & -21.6300 & 1.2046\\
5.88 & -21.7424 & -21.6615 & 1.2047\\
5.92 & -21.7576 & -21.6766 & 1.2049\\
5.96 & -21.7696 & -21.6886 & 1.2050\\
\end{tabular}
\end{table}  
\begin{table}[!htbp]
\begin{tabular}{c  c c c}  
log~$T$& log $\Lambda_N$ & log $\Lambda_{\rm hd}$ & $n_{\rm e}/n_{\rm H}$\\
(K) & (erg~s$^{-1}$~cm$^{3}$) & (erg~s$^{-1}$~cm$^{3}$) & \\
\hline
6.00 & -21.7883 & -21.7073 & 1.2051\\
6.04 & -21.8115 & -21.7304 & 1.2053\\
6.08 & -21.8303 & -21.7491 & 1.2055\\
6.12 & -21.8419 & -21.7607 & 1.2056\\
6.16 & -21.8514 & -21.7701 & 1.2058\\
6.20 & -21.8690 & -21.7877 & 1.2060\\
6.24 & -21.9057 & -21.8243 & 1.2062\\
6.28 & -21.9690 & -21.8875 & 1.2065\\
6.32 & -22.0554 & -21.9738 & 1.2067\\
6.36 & -22.1488 & -22.0671 & 1.2070\\
6.40 & -22.2355 & -22.1537 & 1.2072\\
6.44 & -22.3084 & -22.2265 & 1.2075\\
6.48 & -22.3641 & -22.2821 & 1.2077\\
6.52 & -22.4033 & -22.3213 & 1.2078\\
6.56 & -22.4282 & -22.3462 & 1.2079\\
6.60 & -22.4408 & -22.3587 & 1.2080\\
6.64 & -22.4443 & -22.3622 & 1.2081\\
6.68 & -22.4411 & -22.3590 & 1.2082\\
6.72 & -22.4334 & -22.3512 & 1.2083\\
6.76 & -22.4242 & -22.3420 & 1.2083\\
6.80 & -22.4164 & -22.3342 & 1.2084\\
6.84 & -22.4134 & -22.3312 & 1.2084\\
6.88 & -22.4168 & -22.3346 & 1.2085\\
6.92 & -22.4267 & -22.3445 & 1.2085\\
6.96 & -22.4418 & -22.3595 & 1.2086\\
7.00 & -22.4603 & -22.3780 & 1.2086\\
7.04 & -22.4830 & -22.4007 & 1.2087\\
7.08 & -22.5112 & -22.4289 & 1.2087\\
7.12 & -22.5449 & -22.4625 & 1.2088\\
7.16 & -22.5819 & -22.4995 & 1.2088\\
7.20 & -22.6177 & -22.5353 & 1.2089\\
7.24 & -22.6483 & -22.5659 & 1.2089\\
7.28 & -22.6719 & -22.5895 & 1.2089\\
7.32 & -22.6883 & -22.6059 & 1.2089\\
7.36 & -22.6985 & -22.6161 & 1.2089\\
7.40 & -22.7032 & -22.6208 & 1.2090\\
7.44 & -22.7037 & -22.6213 & 1.2090\\
7.48 & -22.7008 & -22.6184 & 1.2090\\
7.52 & -22.6950 & -22.6126 & 1.2090\\
7.56 & -22.6869 & -22.6045 & 1.2090\\
7.60 & -22.6769 & -22.5945 & 1.2090\\
7.64 & -22.6655 & -22.5831 & 1.2090\\
7.68 & -22.6531 & -22.5707 & 1.2090\\
7.72 & -22.6397 & -22.5573 & 1.2090\\
7.76 & -22.6258 & -22.5434 & 1.2090\\
7.80 & -22.6111 & -22.5287 & 1.2090\\
7.84 & -22.5964 & -22.5140 & 1.2090\\
7.88 & -22.5816 & -22.4992 & 1.2090\\
7.92 & -22.5668 & -22.4844 & 1.2090\\
7.96 & -22.5519 & -22.4695 & 1.2090\\
8.00 & -22.5367 & -22.4543 & 1.2090\\
8.04 & -22.5216 & -22.4392 & 1.2090\\
8.08 & -22.5062 & -22.4237 & 1.2091\\
8.12 & -22.4912 & -22.4087 & 1.2091\\
8.16 & -22.4753 & -22.3928 & 1.2091\\
\hline                                   
\end{tabular}
\end{table}  

 \begin{acknowledgements}
We thank the referee, J.C. Raymond, for his helpful comments that helped us to improve the manuscript.
This study has been financially supported by a Vidi grant from the Netherlands Organisation for Scientific Research (NWO). This work was sponsored by the Stichting Nationale Computerfaciliteiten (National Computing Facilities Foundation, NCF) for the use of supercomputer facilities, with financial support from the Nederlandse Organisatie voor Wetenschappelijk Onderzoek (Netherlands Organisation for Scientific Research, NWO). 
 \end{acknowledgements}

\bibliography{../adssample}

\clearpage

\begin{table}[!htbp]
\centering                 
\caption{\textrm{Cooling curve for solar metallicity for $T<10^4$~K}, as computed from \citet{1972DalgarnoMcCray}, see Sect.~\ref{sec:DMcooling}.}           
\label{table:DMsolar}        
\begin{tabular}{c  c c c c}       
\hline\hline                    
log~$T$& $\log \Lambda_{hd}$ &$\log \Lambda_{hd}$ &$\log \Lambda_{hd}$ & $\log \Lambda_{hd}$ \\
 &$f_i=10^{-4}$ & $f_i=10^{-3}$ &$f_i=10^{-2}$ & $f_i=10^{-1}$\\
\hline
 1.00  & -31.0377  & -30.0377  & -29.0377  & -28.0377\\
 1.04  & -30.7062  & -29.7062  & -28.7062  & -27.7062\\
 1.08  & -30.4055  & -29.4055  & -28.4055  & -27.4055\\
 1.12  & -30.1331  & -29.1331  & -28.1331  & -27.1331\\
 1.16  & -29.8864  & -28.8864  & -27.8864  & -26.8864\\
 1.20  & -29.6631  & -28.6631  & -27.6631  & -26.6631\\
 1.24  & -29.4614  & -28.4614  & -27.4614  & -26.4614\\
 1.28  & -29.2791  & -28.2791  & -27.2791  & -26.2791\\
 1.32  & -29.1146  & -28.1146  & -27.1146  & -26.1146\\
 1.36  & -28.9662  & -27.9662  & -26.9662  & -25.9662\\
 1.40  & -28.8330  & -27.8330  & -26.8330  & -25.8330\\
 1.44  & -28.7129  & -27.7129  & -26.7129  & -25.7129\\
 1.48  & -28.6052  & -27.6052  & -26.6052  & -25.6052\\
 1.52  & -28.5086  & -27.5088  & -26.5088  & -25.5088\\
 1.56  & -28.4222  & -27.4225  & -26.4225  & -25.4225\\
 1.60  & -28.3447  & -27.3454  & -26.3455  & -25.3455\\
 1.64  & -28.2751  & -27.2767  & -26.2769  & -25.2769\\
 1.68  & -28.2120  & -27.2153  & -26.2157  & -25.2157\\
 1.72  & -28.1541  & -27.1605  & -26.1611  & -25.1612\\
 1.76  & -28.0995  & -27.1111  & -26.1123  & -25.1124\\
 1.80  & -28.0460  & -27.0664  & -26.0684  & -25.0686\\
 1.84  & -27.9914  & -27.0251  & -26.0286  & -25.0290\\
 1.88  & -27.9333  & -26.9863  & -25.9918  & -24.9927\\
 1.92  & -27.8697  & -26.9488  & -25.9578  & -24.9586\\
 1.96  & -27.7989  & -26.9119  & -25.9248  & -24.9263\\
 2.00  & -27.7206  & -26.8742  & -25.8931  & -24.8948\\
 2.04  & -27.6353  & -26.8353  & -25.8614  & -24.8642\\
 2.08  & -27.5447  & -26.7948  & -25.8300  & -24.8336\\
 2.12  & -27.4506  & -26.7523  & -25.7983  & -24.8030\\
 2.16  & -27.3551  & -26.7080  & -25.7660  & -24.7724\\
 2.20  & -27.2597  & -26.6619  & -25.7338  & -24.7416\\
 2.24  & -27.1661  & -26.6146  & -25.7011  & -24.7109\\
 2.28  & -27.0751  & -26.5666  & -25.6690  & -24.6807\\
 2.32  & -26.9876  & -26.5183  & -25.6370  & -24.6509\\
 2.36  & -26.9041  & -26.4702  & -25.6057  & -24.6220\\
 2.40  & -26.8245  & -26.4229  & -25.5754  & -24.5941\\
 2.44  & -26.7496  & -26.3765  & -25.5464  & -24.5677\\
 2.48  & -26.6788  & -26.3317  & -25.5186  & -24.5426\\
 2.52  & -26.6124  & -26.2886  & -25.4923  & -24.5190\\
 2.56  & -26.5504  & -26.2473  & -25.4674  & -24.4970\\
 2.60  & -26.4924  & -26.2078  & -25.4439  & -24.4765\\
 2.64  & -26.4383  & -26.1704  & -25.4219  & -24.4574\\
 2.68  & -26.3880  & -26.1348  & -25.4011  & -24.4395\\
 2.72  & -26.3412  & -26.1012  & -25.3813  & -24.4226\\
 2.76  & -26.2978  & -26.0692  & -25.3623  & -24.4063\\
 2.80  & -26.2576  & -26.0389  & -25.3437  & -24.3903\\
 2.84  & -26.2203  & -26.0101  & -25.3254  & -24.3744\\
 2.88  & -26.1859  & -25.9825  & -25.3071  & -24.3582\\
 2.92  & -26.1540  & -25.9566  & -25.2888  & -24.3418\\
 2.96  & -26.1246  & -25.9318  & -25.2703  & -24.3249\\
\end{tabular}
\end{table}  
\begin{table}[!htbp]
\centering                 
\begin{tabular}{c  c c c c}       
log~$T$& $\log \Lambda_{hd}$ &$\log \Lambda_{hd}$ &$\log \Lambda_{hd}$ & $\log \Lambda_{hd}$ \\
 &$f_i=10^{-4}$ & $f_i=10^{-3}$ &$f_i=10^{-2}$ & $f_i=10^{-1}$\\
\hline
 3.00  & -26.0975  & -25.9083  & -25.2517  & -24.3076\\
 3.04  & -26.0724  & -25.8857  & -25.2332  & -24.2901\\
 3.08  & -26.0493  & -25.8645  & -25.2149  & -24.2724\\
 3.12  & -26.0281  & -25.8447  & -25.1969  & -24.2549\\
 3.16  & -26.0085  & -25.8259  & -25.1795  & -24.2378\\
 3.20  & -25.9905  & -25.8085  & -25.1630  & -24.2215\\
 3.24  & -25.9743  & -25.7926  & -25.1474  & -24.2061\\
 3.28  & -25.9590  & -25.7778  & -25.1330  & -24.1918\\
 3.32  & -25.9454  & -25.7642  & -25.1199  & -24.1787\\
 3.36  & -25.9326  & -25.7520  & -25.1081  & -24.1672\\
 3.40  & -25.9212  & -25.7409  & -25.0977  & -24.1570\\
 3.44  & -25.9104  & -25.7310  & -25.0887  & -24.1482\\
 3.48  & -25.9010  & -25.7222  & -25.0809  & -24.1407\\
 3.52  & -25.8925  & -25.7142  & -25.0742  & -24.1342\\
 3.56  & -25.8844  & -25.7071  & -25.0683  & -24.1287\\
 3.60  & -25.8771  & -25.7005  & -25.0627  & -24.1234\\
 3.64  & -25.8703  & -25.6942  & -25.0570  & -24.1178\\
 3.68  & -25.8642  & -25.6878  & -25.0505  & -24.1112\\
 3.72  & -25.8586  & -25.6811  & -25.0422  & -24.1025\\
 3.76  & -25.8529  & -25.6733  & -25.0312  & -24.0907\\
 3.80  & -25.8474  & -25.6641  & -25.0161  & -24.0741\\
 3.84  & -25.8422  & -25.6525  & -24.9957  & -24.0514\\
 3.88  & -25.8356  & -25.6325  & -24.9570  & -24.0081\\
 3.92  & -25.8286  & -25.6080  & -24.9104  & -23.9566\\
 3.96  & -25.8133  & -25.5367  & -24.7799  & -23.8139\\
 4.00  & -25.7997  & -25.4806  & -24.6878  & -23.7151\\
 \hline                                   
\end{tabular}
\end{table}

\begin{sidewaystable*}[!htbp]
\centering                 
\caption{\textrm{Cooling rates per element for solar metallicity} }           
\label{table:KSabun}        
\begin{tabular}{c  c c c c c c c c c c c c c c c}       
\hline\hline               
\tiny
T & H & He & C & N & O & Ne & Na & Mg & Al & Si & S & Ar & Ca & Fe & Ni\\
\hline
4.20 & -21.34 & -25.79 & -23.63 & -24.86 & -24.94 & -26.60 & -29.35 & -27.47 & -24.51 & -23.93 & -26.58 & -28.18 & -28.59 & -27.37 & -28.96\\
4.24 & -21.37 & -25.25 & -23.39 & -24.63 & -24.55 & -26.16 & -29.35 & -27.34 & -24.31 & -23.60 & -26.33 & -28.01 & -28.56 & -27.23 & -28.92\\
4.28 & -21.47 & -24.76 & -23.16 & -24.41 & -24.18 & -25.77 & -29.35 & -27.26 & -24.13 & -23.30 & -26.09 & -27.95 & -28.55 & -27.12 & -28.86\\
4.32 & -21.59 & -24.31 & -22.95 & -24.18 & -23.84 & -25.45 & -29.33 & -27.24 & -23.97 & -23.00 & -25.85 & -27.93 & -28.55 & -27.06 & -28.80\\
4.36 & -21.71 & -23.91 & -22.75 & -23.95 & -23.54 & -25.22 & -29.23 & -27.25 & -23.83 & -22.74 & -25.59 & -27.93 & -28.55 & -27.04 & -28.77\\
4.40 & -21.83 & -23.58 & -22.56 & -23.73 & -23.27 & -25.00 & -28.98 & -27.27 & -23.71 & -22.52 & -25.34 & -27.95 & -28.56 & -27.04 & -28.75\\
4.44 & -21.94 & -23.34 & -22.37 & -23.52 & -23.02 & -25.05 & -28.58 & -27.28 & -23.61 & -22.33 & -25.11 & -27.96 & -28.56 & -27.04 & -28.72\\
4.48 & -22.05 & -23.22 & -22.18 & -23.32 & -22.79 & -25.10 & -28.12 & -27.30 & -23.53 & -22.17 & -24.90 & -27.97 & -28.57 & -27.03 & -28.67\\
4.52 & -22.15 & -23.22 & -22.00 & -23.12 & -22.59 & -25.03 & -27.66 & -27.31 & -23.49 & -22.04 & -24.71 & -27.98 & -28.56 & -27.01 & -28.57\\
4.56 & -22.25 & -23.26 & -21.82 & -22.94 & -22.40 & -24.83 & -27.24 & -27.30 & -23.51 & -21.93 & -24.53 & -27.99 & -28.47 & -26.97 & -28.44\\
4.60 & -22.34 & -23.29 & -21.66 & -22.75 & -22.23 & -24.58 & -26.85 & -27.23 & -23.62 & -21.83 & -24.34 & -28.00 & -28.14 & -26.89 & -28.31\\
4.64 & -22.42 & -23.23 & -21.51 & -22.57 & -22.07 & -24.32 & -26.50 & -27.02 & -23.81 & -21.76 & -24.15 & -28.01 & -27.61 & -26.82 & -28.22\\
4.68 & -22.51 & -23.06 & -21.39 & -22.41 & -21.93 & -24.08 & -26.18 & -26.68 & -24.05 & -21.70 & -23.94 & -28.01 & -27.09 & -26.76 & -28.14\\
4.72 & -22.59 & -22.80 & -21.28 & -22.27 & -21.80 & -23.85 & -25.89 & -26.28 & -24.30 & -21.65 & -23.72 & -28.01 & -26.64 & -26.71 & -28.06\\
4.76 & -22.67 & -22.53 & -21.20 & -22.14 & -21.68 & -23.66 & -25.64 & -25.87 & -24.54 & -21.63 & -23.50 & -28.00 & -26.25 & -26.66 & -27.99\\
4.80 & -22.74 & -22.27 & -21.13 & -22.03 & -21.58 & -23.48 & -25.46 & -25.49 & -24.78 & -21.65 & -23.29 & -27.98 & -25.91 & -26.61 & -27.91\\
4.84 & -22.81 & -22.07 & -21.08 & -21.95 & -21.48 & -23.33 & -25.32 & -25.14 & -24.99 & -21.75 & -23.09 & -27.96 & -25.60 & -26.54 & -27.84\\
4.88 & -22.88 & -21.96 & -21.05 & -21.87 & -21.39 & -23.19 & -25.22 & -24.82 & -25.19 & -21.96 & -22.93 & -27.92 & -25.33 & -26.45 & -27.76\\
4.92 & -22.95 & -21.93 & -21.04 & -21.81 & -21.31 & -23.05 & -25.13 & -24.53 & -25.38 & -22.28 & -22.79 & -27.85 & -25.09 & -26.33 & -27.70\\
4.96 & -23.01 & -21.97 & -21.06 & -21.76 & -21.23 & -22.93 & -25.06 & -24.27 & -25.55 & -22.63 & -22.67 & -27.65 & -24.88 & -26.19 & -27.63\\
5.00 & -23.07 & -22.05 & -21.15 & -21.71 & -21.16 & -22.80 & -24.97 & -24.04 & -25.70 & -22.97 & -22.58 & -27.24 & -24.70 & -26.03 & -27.58\\
5.04 & -23.13 & -22.14 & -21.32 & -21.66 & -21.09 & -22.67 & -24.88 & -23.85 & -25.84 & -23.28 & -22.50 & -26.72 & -24.54 & -25.89 & -27.52\\
5.08 & -23.18 & -22.24 & -21.56 & -21.63 & -21.03 & -22.55 & -24.78 & -23.70 & -25.96 & -23.55 & -22.44 & -26.20 & -24.42 & -25.78 & -27.47\\
5.12 & -23.23 & -22.34 & -21.85 & -21.60 & -20.97 & -22.44 & -24.67 & -23.58 & -26.07 & -23.78 & -22.39 & -25.73 & -24.32 & -25.68 & -27.43\\
5.16 & -23.27 & -22.43 & -22.13 & -21.60 & -20.93 & -22.34 & -24.56 & -23.49 & -26.18 & -23.96 & -22.36 & -25.32 & -24.25 & -25.59 & -27.40\\
5.20 & -23.32 & -22.52 & -22.40 & -21.64 & -20.89 & -22.25 & -24.46 & -23.41 & -26.30 & -24.06 & -22.37 & -24.97 & -24.20 & -25.42 & -27.36\\
5.24 & -23.35 & -22.61 & -22.65 & -21.74 & -20.85 & -22.18 & -24.35 & -23.34 & -26.35 & -24.08 & -22.46 & -24.68 & -24.18 & -25.09 & -27.33\\
5.28 & -23.39 & -22.69 & -22.87 & -21.94 & -20.82 & -22.12 & -24.25 & -23.26 & -26.22 & -24.01 & -22.66 & -24.43 & -24.17 & -24.64 & -27.30\\
5.32 & -23.42 & -22.77 & -23.07 & -22.21 & -20.80 & -22.06 & -24.15 & -23.16 & -25.91 & -23.89 & -22.94 & -24.23 & -24.17 & -24.18 & -27.26\\
5.36 & -23.45 & -22.85 & -23.26 & -22.51 & -20.80 & -22.01 & -24.07 & -23.07 & -25.55 & -23.76 & -23.24 & -24.06 & -24.17 & -23.76 & -27.20\\
5.40 & -23.47 & -22.92 & -23.43 & -22.79 & -20.84 & -21.97 & -24.00 & -22.97 & -25.21 & -23.65 & -23.51 & -23.93 & -24.14 & -23.40 & -27.14\\
5.44 & -23.49 & -22.98 & -23.58 & -23.05 & -20.94 & -21.93 & -23.93 & -22.88 & -24.91 & -23.56 & -23.74 & -23.84 & -24.09 & -23.10 & -27.05\\
5.48 & -23.51 & -23.05 & -23.73 & -23.29 & -21.14 & -21.89 & -23.88 & -22.79 & -24.66 & -23.49 & -23.91 & -23.78 & -24.03 & -22.86 & -26.97\\
5.52 & -23.53 & -23.11 & -23.86 & -23.49 & -21.41 & -21.86 & -23.84 & -22.72 & -24.46 & -23.43 & -24.00 & -23.78 & -23.98 & -22.67 & -26.88\\
5.56 & -23.54 & -23.17 & -23.98 & -23.68 & -21.69 & -21.84 & -23.80 & -22.66 & -24.30 & -23.38 & -24.00 & -23.89 & -23.94 & -22.51 & -26.78\\
5.60 & -23.56 & -23.22 & -24.07 & -23.85 & -21.96 & -21.81 & -23.78 & -22.62 & -24.17 & -23.32 & -23.94 & -24.09 & -23.91 & -22.39 & -26.64\\
5.64 & -23.57 & -23.27 & -24.13 & -24.00 & -22.20 & -21.79 & -23.76 & -22.58 & -24.07 & -23.28 & -23.85 & -24.30 & -23.87 & -22.30 & -26.45\\
5.68 & -23.57 & -23.32 & -24.14 & -24.14 & -22.42 & -21.79 & -23.76 & -22.55 & -23.99 & -23.23 & -23.75 & -24.47 & -23.81 & -22.23 & -26.19\\
5.72 & -23.58 & -23.36 & -24.10 & -24.26 & -22.61 & -21.83 & -23.76 & -22.53 & -23.93 & -23.20 & -23.65 & -24.57 & -23.74 & -22.17 & -25.89\\
5.76 & -23.58 & -23.40 & -24.02 & -24.36 & -22.77 & -21.94 & -23.78 & -22.51 & -23.88 & -23.19 & -23.58 & -24.59 & -23.72 & -22.12 & -25.60\\
5.80 & -23.58 & -23.44 & -23.91 & -24.43 & -22.92 & -22.14 & -23.82 & -22.49 & -23.86 & -23.18 & -23.53 & -24.54 & -23.83 & -22.08 & -25.33\\
5.84 & -23.58 & -23.47 & -23.81 & -24.47 & -23.06 & -22.38 & -23.88 & -22.47 & -23.85 & -23.16 & -23.50 & -24.46 & -24.05 & -22.05 & -25.08\\
5.88 & -23.58 & -23.51 & -23.71 & -24.46 & -23.17 & -22.63 & -23.99 & -22.45 & -23.86 & -23.11 & -23.48 & -24.37 & -24.29 & -22.03 & -24.86\\
5.92 & -23.58 & -23.53 & -23.63 & -24.41 & -23.27 & -22.86 & -24.17 & -22.45 & -23.89 & -23.04 & -23.46 & -24.29 & -24.45 & -22.02 & -24.66\\
5.96 & -23.58 & -23.56 & -23.56 & -24.32 & -23.33 & -23.07 & -24.39 & -22.48 & -23.94 & -22.96 & -23.44 & -24.22 & -24.52 & -22.01 & -24.48\\

\end{tabular}
\end{sidewaystable*}  
\begin{sidewaystable*}[!htbp]
\centering                 
\begin{tabular}{c  c c c c c c c c c c c c c c c}       
T & H & He & C & N & O & Ne & Na & Mg & Al & Si & S & Ar & Ca & Fe & Ni\\
\hline
6.00 & -23.57 & -23.58 & -23.52 & -24.23 & -23.37 & -23.26 & -24.61 & -22.57 & -24.00 & -22.88 & -23.42 & -24.17 & -24.51 & -22.01 & -24.32\\
6.04 & -23.57 & -23.60 & -23.49 & -24.14 & -23.36 & -23.42 & -24.82 & -22.74 & -24.05 & -22.82 & -23.38 & -24.15 & -24.46 & -22.01 & -24.17\\
6.08 & -23.56 & -23.62 & -23.50 & -24.05 & -23.33 & -23.56 & -25.01 & -22.95 & -24.11 & -22.77 & -23.35 & -24.14 & -24.40 & -22.02 & -24.04\\
6.12 & -23.55 & -23.64 & -23.53 & -23.98 & -23.26 & -23.68 & -25.18 & -23.17 & -24.19 & -22.74 & -23.32 & -24.15 & -24.33 & -22.02 & -23.92\\
6.16 & -23.54 & -23.65 & -23.58 & -23.93 & -23.19 & -23.79 & -25.33 & -23.37 & -24.31 & -22.73 & -23.30 & -24.17 & -24.28 & -22.03 & -23.82\\
6.20 & -23.53 & -23.66 & -23.64 & -23.90 & -23.11 & -23.86 & -25.47 & -23.56 & -24.47 & -22.77 & -23.29 & -24.20 & -24.25 & -22.05 & -23.74\\
6.24 & -23.52 & -23.67 & -23.71 & -23.89 & -23.04 & -23.91 & -25.59 & -23.72 & -24.64 & -22.87 & -23.29 & -24.24 & -24.24 & -22.09 & -23.67\\
6.28 & -23.51 & -23.68 & -23.79 & -23.91 & -22.98 & -23.94 & -25.68 & -23.86 & -24.81 & -23.02 & -23.29 & -24.28 & -24.23 & -22.17 & -23.63\\
6.32 & -23.50 & -23.68 & -23.86 & -23.95 & -22.93 & -23.93 & -25.75 & -23.98 & -24.96 & -23.19 & -23.29 & -24.31 & -24.23 & -22.29 & -23.61\\
6.36 & -23.49 & -23.68 & -23.93 & -24.01 & -22.91 & -23.90 & -25.78 & -24.08 & -25.11 & -23.36 & -23.32 & -24.32 & -24.22 & -22.44 & -23.63\\
6.40 & -23.47 & -23.69 & -24.00 & -24.07 & -22.91 & -23.86 & -25.78 & -24.16 & -25.23 & -23.52 & -23.39 & -24.32 & -24.19 & -22.58 & -23.72\\
6.44 & -23.46 & -23.69 & -24.06 & -24.14 & -22.94 & -23.80 & -25.76 & -24.22 & -25.33 & -23.67 & -23.50 & -24.31 & -24.16 & -22.69 & -23.87\\
6.48 & -23.45 & -23.68 & -24.12 & -24.21 & -22.98 & -23.75 & -25.71 & -24.25 & -25.41 & -23.79 & -23.63 & -24.31 & -24.14 & -22.76 & -23.98\\
6.52 & -23.43 & -23.68 & -24.18 & -24.28 & -23.04 & -23.70 & -25.65 & -24.26 & -25.47 & -23.90 & -23.76 & -24.35 & -24.12 & -22.80 & -24.04\\
6.56 & -23.42 & -23.68 & -24.23 & -24.34 & -23.10 & -23.66 & -25.58 & -24.24 & -25.49 & -23.98 & -23.89 & -24.43 & -24.12 & -22.81 & -24.05\\
6.60 & -23.40 & -23.67 & -24.28 & -24.41 & -23.17 & -23.64 & -25.52 & -24.22 & -25.49 & -24.05 & -24.00 & -24.55 & -24.14 & -22.80 & -24.05\\
6.64 & -23.39 & -23.67 & -24.32 & -24.46 & -23.23 & -23.63 & -25.47 & -24.18 & -25.46 & -24.09 & -24.10 & -24.68 & -24.20 & -22.78 & -24.04\\
6.68 & -23.37 & -23.66 & -24.36 & -24.52 & -23.30 & -23.64 & -25.42 & -24.14 & -25.42 & -24.12 & -24.19 & -24.81 & -24.29 & -22.74 & -24.02\\
6.72 & -23.36 & -23.65 & -24.40 & -24.57 & -23.36 & -23.66 & -25.39 & -24.10 & -25.37 & -24.13 & -24.26 & -24.91 & -24.41 & -22.71 & -24.02\\
6.76 & -23.34 & -23.64 & -24.44 & -24.61 & -23.41 & -23.70 & -25.38 & -24.07 & -25.31 & -24.12 & -24.31 & -25.00 & -24.55 & -22.68 & -24.03\\
6.80 & -23.33 & -23.64 & -24.47 & -24.66 & -23.47 & -23.75 & -25.39 & -24.05 & -25.26 & -24.10 & -24.34 & -25.06 & -24.67 & -22.65 & -24.06\\
6.84 & -23.31 & -23.63 & -24.50 & -24.70 & -23.52 & -23.80 & -25.41 & -24.03 & -25.21 & -24.07 & -24.36 & -25.11 & -24.79 & -22.64 & -24.10\\
6.88 & -23.29 & -23.61 & -24.53 & -24.73 & -23.56 & -23.86 & -25.44 & -24.04 & -25.17 & -24.05 & -24.37 & -25.14 & -24.89 & -22.64 & -24.15\\
6.92 & -23.28 & -23.60 & -24.55 & -24.77 & -23.61 & -23.92 & -25.48 & -24.05 & -25.15 & -24.02 & -24.36 & -25.15 & -24.97 & -22.65 & -24.20\\
6.96 & -23.26 & -23.59 & -24.57 & -24.80 & -23.65 & -23.97 & -25.53 & -24.08 & -25.14 & -24.00 & -24.35 & -25.15 & -25.04 & -22.67 & -24.25\\
7.00 & -23.24 & -23.58 & -24.59 & -24.82 & -23.68 & -24.02 & -25.58 & -24.12 & -25.14 & -23.98 & -24.33 & -25.13 & -25.09 & -22.70 & -24.29\\
7.04 & -23.23 & -23.57 & -24.61 & -24.85 & -23.72 & -24.07 & -25.63 & -24.16 & -25.15 & -23.98 & -24.30 & -25.11 & -25.12 & -22.74 & -24.32\\
7.08 & -23.21 & -23.55 & -24.62 & -24.87 & -23.75 & -24.12 & -25.68 & -24.21 & -25.18 & -23.98 & -24.28 & -25.07 & -25.14 & -22.79 & -24.35\\
7.12 & -23.19 & -23.54 & -24.63 & -24.89 & -23.77 & -24.16 & -25.73 & -24.26 & -25.22 & -24.00 & -24.26 & -25.04 & -25.15 & -22.86 & -24.39\\
7.16 & -23.17 & -23.52 & -24.64 & -24.91 & -23.80 & -24.20 & -25.77 & -24.30 & -25.26 & -24.02 & -24.25 & -25.00 & -25.15 & -22.94 & -24.45\\
7.20 & -23.16 & -23.51 & -24.65 & -24.92 & -23.82 & -24.23 & -25.81 & -24.35 & -25.30 & -24.06 & -24.24 & -24.96 & -25.13 & -23.03 & -24.51\\
7.24 & -23.14 & -23.50 & -24.66 & -24.94 & -23.84 & -24.27 & -25.85 & -24.40 & -25.34 & -24.09 & -24.24 & -24.93 & -25.11 & -23.13 & -24.58\\
7.28 & -23.12 & -23.48 & -24.66 & -24.95 & -23.86 & -24.30 & -25.89 & -24.44 & -25.39 & -24.13 & -24.25 & -24.90 & -25.09 & -23.21 & -24.64\\
7.32 & -23.10 & -23.46 & -24.67 & -24.96 & -23.87 & -24.32 & -25.92 & -24.47 & -25.43 & -24.18 & -24.27 & -24.88 & -25.07 & -23.29 & -24.70\\
7.36 & -23.08 & -23.45 & -24.67 & -24.97 & -23.88 & -24.35 & -25.95 & -24.51 & -25.47 & -24.22 & -24.30 & -24.87 & -25.05 & -23.36 & -24.75\\
7.40 & -23.07 & -23.43 & -24.67 & -24.97 & -23.90 & -24.37 & -25.97 & -24.54 & -25.50 & -24.26 & -24.33 & -24.86 & -25.03 & -23.42 & -24.80\\
7.44 & -23.05 & -23.42 & -24.67 & -24.98 & -23.90 & -24.39 & -26.00 & -24.57 & -25.54 & -24.29 & -24.36 & -24.87 & -25.01 & -23.46 & -24.84\\
7.48 & -23.03 & -23.40 & -24.67 & -24.98 & -23.91 & -24.41 & -26.02 & -24.60 & -25.57 & -24.33 & -24.40 & -24.88 & -25.00 & -23.49 & -24.87\\
7.52 & -23.01 & -23.38 & -24.66 & -24.98 & -23.92 & -24.42 & -26.04 & -24.62 & -25.60 & -24.36 & -24.43 & -24.90 & -25.00 & -23.52 & -24.90\\
7.56 & -22.99 & -23.37 & -24.66 & -24.98 & -23.92 & -24.44 & -26.05 & -24.64 & -25.62 & -24.39 & -24.47 & -24.92 & -25.00 & -23.53 & -24.92\\
7.60 & -22.97 & -23.35 & -24.65 & -24.98 & -23.92 & -24.45 & -26.07 & -24.66 & -25.65 & -24.42 & -24.50 & -24.95 & -25.01 & -23.54 & -24.93\\
7.64 & -22.95 & -23.33 & -24.65 & -24.97 & -23.93 & -24.45 & -26.08 & -24.68 & -25.67 & -24.45 & -24.53 & -24.98 & -25.02 & -23.54 & -24.94\\
7.68 & -22.94 & -23.32 & -24.64 & -24.97 & -23.92 & -24.46 & -26.09 & -24.70 & -25.68 & -24.47 & -24.56 & -25.01 & -25.04 & -23.54 & -24.94\\
7.72 & -22.92 & -23.30 & -24.63 & -24.97 & -23.92 & -24.47 & -26.10 & -24.71 & -25.70 & -24.49 & -24.59 & -25.03 & -25.06 & -23.54 & -24.94\\
7.76 & -22.90 & -23.28 & -24.62 & -24.96 & -23.92 & -24.47 & -26.10 & -24.72 & -25.71 & -24.51 & -24.61 & -25.06 & -25.08 & -23.53 & -24.93\\
7.80 & -22.88 & -23.26 & -24.61 & -24.95 & -23.92 & -24.47 & -26.11 & -24.73 & -25.72 & -24.52 & -24.64 & -25.09 & -25.10 & -23.53 & -24.93\\
7.84 & -22.86 & -23.25 & -24.60 & -24.95 & -23.91 & -24.47 & -26.11 & -24.73 & -25.73 & -24.53 & -24.66 & -25.11 & -25.12 & -23.53 & -24.92\\
7.88 & -22.84 & -23.23 & -24.59 & -24.94 & -23.91 & -24.47 & -26.11 & -24.74 & -25.74 & -24.55 & -24.67 & -25.13 & -25.15 & -23.52 & -24.91\\
7.92 & -22.82 & -23.21 & -24.58 & -24.93 & -23.90 & -24.47 & -26.12 & -24.74 & -25.75 & -24.56 & -24.69 & -25.15 & -25.17 & -23.52 & -24.90\\
7.96 & -22.80 & -23.19 & -24.57 & -24.92 & -23.89 & -24.47 & -26.11 & -24.75 & -25.75 & -24.56 & -24.70 & -25.17 & -25.19 & -23.53 & -24.89\\
8.00 & -22.78 & -23.17 & -24.56 & -24.91 & -23.88 & -24.46 & -26.11 & -24.75 & -25.76 & -24.57 & -24.71 & -25.19 & -25.21 & -23.53 & -24.89\\
8.04 & -22.76 & -23.16 & -24.54 & -24.90 & -23.87 & -24.46 & -26.11 & -24.75 & -25.76 & -24.57 & -24.72 & -25.20 & -25.22 & -23.54 & -24.89\\
8.08 & -22.74 & -23.14 & -24.53 & -24.88 & -23.86 & -24.45 & -26.10 & -24.74 & -25.76 & -24.57 & -24.73 & -25.21 & -25.24 & -23.55 & -24.89\\
8.12 & -22.72 & -23.12 & -24.52 & -24.87 & -23.85 & -24.45 & -26.10 & -24.74 & -25.75 & -24.58 & -24.73 & -25.22 & -25.25 & -23.56 & -24.89\\
8.16 & -22.70 & -23.10 & -24.50 & -24.86 & -23.84 & -24.44 & -26.09 & -24.74 & -25.75 & -24.58 & -24.74 & -25.23 & -25.27 & -23.57 & -24.89\\
\hline                                   
\end{tabular}
\end{sidewaystable*}  
\end{document}